\documentstyle[aps]{revtex}
\newtheorem{teo}{Theorem}

\begin{document}

\makeatletter \title{\bf Magnetic monopoles in $U(1)_{4}$ Lattice Gauge
Theory with Wilson action}
\author{V. Cirigliano $^{a,b}$  and  G. Paffuti $^{a,c}$}
\address{$^{a}$ Dipartimento di Fisica dell'Universit\`a and I.N.F.N., P.zza
Torricelli 2, I-56100 Pisa, Italy\\ $^{b}$ {\rm E-mail:
ciriglia@ipifidpt.difi.unipi.it } \\ $^{c}$ {\rm E-mail:
paffuti@ipifidpt.difi.unipi.it }}
\makeatother \maketitle
\begin{abstract}
\\ We construct the Euclidean Green functions for the soliton (magnetic
monopole) field in the $U(1)_{4}$ Lattice Gauge Theory with Wilson
action. We show that in the strong coupling regime there is monopole
 condensation while in the QED phase the physical Hilbert space splits into
orthogonal soliton sectors labelled by integer magnetic charge.
\vspace{.2cm}\\
Preprint IFUP-TH/29-97
\end{abstract}

\tableofcontents

\section{Introduction}
In this paper we apply the methods introduced in \cite{FM2} to the
construction of soliton (magnetic monopole) sectors for the $U(1)$
Lattice Gauge Theory with Wilson action \cite{Wils}. The soliton
quantization for the $U(1)$ Lattice Gauge Theory with Villain action
\cite{Vill}, as well as for a large class of models, has been carried
out in \cite{FM2,FM1} by constructing the Euclidean Green functions of
soliton fields as expectation values of suitable disorder operators.
These operators are obtained by coupling the theory to a generalized
external gauge field in an hyper-gauge invariant way, as we shall
briefly recall below. In the statistical approach this procedure
corresponds to the introduction of open-ended line defects.  In the
case of the $U(1)_{4}$ gauge theories such line defects are just
magnetic loops carrying a defect (topological) charge.  Opening up a
loop one introduces magnetic monopoles at the endpoints of the current
line.  An Osterwalder-Schrader (O.S.) reconstruction theorem
\cite{OS1,OS2} applied to the disorder fields correlation functions
permits the identification of vacuum expectation values of the soliton
field, which can be considered as a charged operator creating magnetic
monopoles \cite{thooft,pisa1}. \\ We have focused our attention in
particular on the vacuum expectation value of the soliton field, which
is defined by a limiting procedure starting from the two point
function:
$$ S_{1} = \lim_{|x| \rightarrow \infty} \, S_{2} (0,x) \ .$$ The
result of our work is that $S_{1}$ is a good disorder parameter for
the phase transition occurring in the model \cite{Guth}. In fact we
show that $S_{1}$ is bounded away from zero in the strong coupling
regime ($\beta \ll 1$) while is vanishing in the QED phase. The proof
of this statement makes use of two different techniques. The strong
coupling phase is analysed by means of a convergent Mayer expansion,
applied to a polymer system obtained by the dual representation
of the model \cite{KennKing}. Besides, the clustering property in the
QED phase is proved using an adapted version of the expansion in
renormalized monopole loops originally given by Fr\"ohlich and Spencer
\cite{FS1,FS2}. The dual representation of the Wilson model has
a measure given by products of modified Bessel functions: the
extimates have been done applying suitable bounds on modified Bessel
functions for $\beta \ll 1$, while using for $\beta$ large an
interpolation to Bessel functions given in \cite{FS1}.\\ The paper is
organised as follows: in this section we shortly define the disorder
fields and correlation functions for the $U(1)$ Wilson model and
describe their connection with magnetic monopoles. Then we enunciate
the reconstruction theorem for the $U(1)$ Wilson model. In section
\ref{condensation} we'll map the model into a polymer system in order
to prove that $S_{1}$ is nonvanishing in the strong coupling regime
and thus the lattice solitons ``condense'' in the vacuum sector.  In
section \ref{sectors} we'll give the expansion in renormalized
monopole loops in order to show that $S_{1} = 0$ in the weak coupling
regime: the Hilbert space of the reconstructed Lattice Quantum Field
Theory splits into orthogonal sectors labelled by the magnetic charge.
Finally, in section \ref{conclusion} we give some concluding remarks.

\subsection{Disorder fields and magnetic monopoles}
In what follows we consider all fields as defined on a finite
lattice $\Lambda \subset Z^{4}$; all estimates needed to proof our
statements are uniform in $|\Lambda |$ and therefore extend to the
thermodynamic limit ($\Lambda \rightarrow Z^{4}$).
The partition function for the $U(1)$ gauge model is
\begin{equation}
\left. Z = \int \, {\cal D} \, \theta \,  \prod_{p \subset
\Lambda} \, \varphi_{\beta} \, ( d \theta_{p} ) \ , \right.
\label{def1}
\end{equation}
where ${\cal D} \theta$ is the product measure on the 1-form $\theta$
 valued in $[- \pi, \pi)$, $d \theta$ is the field strength defined on
 the plaquette $p$ and (for Wilson action)
\begin{equation}
 \left.  \varphi_{\beta} \, (d \theta ) = e^{ \beta \, \cos \left( d \theta
\right) }  \right. \ .
\end{equation}
To define a disorder operator we can consider a modified partition function
in which an external hyper-gauge field strength $X$ is coupled to the dynamical
variables:
\begin{equation}
\left. Z (X) = \int \, {\cal D} \, \theta \,  \prod_{p \subset
\Lambda} \, \varphi_{\beta} \, ( d \theta_{p} + X_{p} )  \ . \right.
\label{def2}
\end{equation}
The mean value of the disorder operator is defined by
\begin{equation}
\left. \langle {\cal D} (X) \rangle = \frac{Z(X)}{Z(0)} \right.
\label{def3}
\end{equation}
and is invariant under the hyper-gauge transformation
\begin{equation}
X \longrightarrow X + d \gamma ,
\label{hyper}
\end{equation}
with  $\gamma$ a generic 1-form. The hyper-gauge invariance follows by
the redefinition of link variables ($\theta \rightarrow  \theta - \gamma$)
 and amounts to say that $\langle {\cal D} (X) \rangle$ depends only on the
3-form $dX=M$. In fact, by Hodge decomposition, on a convex lattice
 one can write
\begin{equation}
X = d \alpha + \delta \frac{1}{\Delta} M
\label{def4}
\end{equation}
and the first term on the right-hand side can be always absorbed in a
 redefinition of $\theta$.  Moreover, $M$ turns out to be the dual of
 the magnetic current density $J^{M}$, because the total field
 strength $G = d\theta + X$ satisfies the following modified Maxwell
 equations (in presence of an electric current $J^{E}$):
\begin{eqnarray}
\delta \, G &  = & J^{E} \\
d \, G & = & M \ \ \ \mbox{or}  \ \ \delta \, * G = J^{M} \ \ ,
\label{def5}
\end{eqnarray}
where by $*G$ we mean the Hodge dual of the form $G$. Finally,
from (\ref{def5}) follows that the magnetic current is identically
conserved
\begin{equation}
\delta \, J^{M} = 0 \ \ .
\label{def6}
\end{equation}
Hence we have that a disorder operator as defined by (\ref{def3}) is
 unavoidably connected to a magnetic current. In this language a
 disorder operator describing an open-ended line defect in the
 statistical system (\ref{def1}) will correspond to a current $J^{M}$
 describing the birth and the evolution of a magnetic monopole.  It is
 possible to parametrize such a conserved magnetic current in a
 general way as follows:
\begin{equation}
J^{M}  = 2 \pi  \, \left[ D - \Omega  \right] \ .
\label{def7}
\end{equation}
Here $\Omega$ is an integer valued 1-form with support on a line
(the Dirac string) whose endpoints are the space-time locations of
monopoles:
\begin{equation}
\left.  \delta \Omega \, (x) = \sum_{i} \, q_{i} \, \delta (x - x_{i})
\ \ \ \ \ \ \ \ \ \ \ \ \ q_{i} \in Z \backslash \{ 0 \} \ \ .
\right.
\label{def7.5}
\end{equation}
$D_{\mu} = (0,\vec{D})$ is a superposition of
Coulomb-like magnetic fields with flux $\pm q_{i}$, spreading out at the
 time slices where the monopoles are located ( $x_{i}^{0}$ )
\begin{equation}
\left. \vec{D} \, (x_{0},\vec{x}) = \pm \frac{q_{i}}{4 \pi} \,
\frac{\vec{x} - \vec{x}_{i}}{ |\vec{x} - \vec{x}_{i}|^{3}} \, \delta (x^{0}
- x_{i}^{0})  \right.
\label{def8}
\end{equation}
in such a way that $\delta D = \delta \Omega$.  Moreover, hyper-gauge
invariance joint to the compactness of the action implies that the
disorder operator does not depend on the shape of the string.  In
conclusion we see that a magnetic monopole of charge $q$ can be
implemented as a defect in the model (\ref{def1}) by a 2-form $X$
whose curvature $dX$ plays the role of a dual magnetic current. For a
 monopole-antimonopole pair of charge $\pm q$ we have
\begin{equation}
\left. * J^{M} =  d \, X = 2 \pi q \, (*D - *\Omega)  \ \ \ \ \ \ \
X = 2 \pi q \, \delta \, \frac{1}{\Delta} \, (*D - *\Omega) \ , \right.
\label{def9}
\end{equation}
where $*\Omega$ is an integer valued 3-form and $*D$ is the
3-form representing the Coulomb field.

\subsection{Mixed order-disorder correlation functions}
Performing a Fourier analysis on (\ref{def2}) we obtain
\begin{equation}
\left.  Z \, (X) =  \sum_{n \, : \, \delta n = 0} \ \prod_{ p \subset
 \Lambda}  I_{\beta} \, (n_{p})  \ \  e^{i (n,X)} \ \ . \right.
\label{mix1}
\end{equation}
$n_{p}$ is the integer valued 2-form labelling the Fourier coefficients.
With $I_{\beta} \, (n)$ we indicate the modified Bessel functions of order
$n$ evaluated in $\beta$ (commonly written as $I_{n}(\beta)$).
On the dual form $v=*n$ the constraint $\delta \, n = 0$ becomes
$d \, v = 0$ and so  we can write $v = d \, A$ and sum over equivalence
classes of integer valued 1-forms defined by $[A] = \{ A' \  : d \, A' =
d \, A \} $:
\begin{equation}
\left.  Z \, (D, \Omega) =  \sum_{[A]} \ \prod_{ p \subset
 \Lambda}  I_{\beta} \, (dA_{p}) \ \  e^{i 2 \pi q (A,D - \Omega)}
 \ \ . \right.
\label{mix2}
\end{equation}
Now $e^{i 2 \pi q (A, \Omega)} = 1$, for integer values of $q$, and we
can conclude that $\langle {\cal D} (D,\Omega) \rangle $ actually
depends only on $(x_{i},q_{i})$, once we have fixed the shape of the
magnetic field $D$ satisfying $\delta \, D = - \sum_{i} \, q_{i} \,
\delta (x - x_{i})$.\\ Now one can introduce ordinary fields,
preserving hypergauge invariance of expectation values:
\begin{equation}
\left.  \psi_{p} (D,\Omega) = e^{i \left[ d \theta_{p} + X_{p} \right]}
 \ . \right.
\label{mix3}
\end{equation}
The correlation functions to which the reconstruction theorem applies are
then given by
\begin{equation}
\left.  S_{n,m} \, (x_{1} \, q_{1},......,x_{n} \, q_{n};\,
p_{1}......p_{m}) = \langle {\cal D} (x_{1} \, q_{1},......,x_{n} \,
q_{n}) \, \psi_{p_{1}}.......  \psi_{p_{m}} \rangle \right.
\label{mix4}
\end{equation}
for $\sum \, q_{i} = 0$. Correlation functions with non vanishing total
charge are defined by the following limiting procedure for $\sum \, q_{i}
 = q$:
\begin{equation}
\left.  S_{n,m} = \lim_{x \rightarrow \infty} \, c_{q} \, S_{n+1,m} \,
 (x_{1},q_{1};......;x_{n},q_{n}; x ,-q; \, p_{1}......p_{m})
\right.
\label{mix5}
\end{equation}
where $c_{q}$ is a normalization constant. We recall now the reconstruction
 theorem in the version given in  \cite{FM2}.
\begin{teo}
If the set of correlation functions $\left\{ S_{n,m} \right\}$ is \\
1) lattice translation invariant;\\
2) O.S. (reflection) positive;\\
3) satisfies cluster properties;\\
then one can reconstruct from  $\left\{ S_{n,m} \right\}$ \\
a) a separeble Hilbert space ${\cal H}$ of physical states;\\
b) a vector $\Omega$ of unit norm, the vacuum;\\
c) a selfadjoint transfer matrix with norm $\|T\| \leq 1$ and unitary
spatial translation operators $U_{\mu} \  \ \mu = 1,..,d-1$ such that
$$ T \, \Omega = U \, \Omega = \Omega \ ; $$
d) $\Omega$ is the unique vector in ${\cal H}$ invariant under $T$ and $U$.\\
If moreover the limits (\ref{mix5}) vanish, then ${\cal H}$ splits into
orthogonal sectors ${\cal H}_{q}  , q \in Z$, which are the lattice
monopole sectors.
\label{teoric}
\end{teo}
In our case hypotheses 1),2),3) follow from traslation invariance and
reflection positivity of the measure defined in the standard way from
Wilson action \cite{Seil}. In particoular one can easily check the
reflection positivity of monomials of disorder fields in the dual
representation, where they assume a standard form and have support on
fixed time slices (because the magnetic field spreads out in fixed
time planes).\\ In the quantum mechanical framework $S_{2} \,
(x,q;y,-q)$ is the amplitude for the creation of a monopole of charge
$q$ at the euclidean point $x$ and its annihilation in $y$. We are
now going to show that in the confining phase the two point function
$S_{2} \, (x,q;y,-q)$ is uniformely bounded away from zero, while in
the QED phase it vanishes at large euclidean time distances:
\begin{equation}
\lim_{|x-y|\rightarrow \infty} \, S_{2} \, (x,q \, ; \, y,-q) = 0 \ .
\label{cluster}
\end{equation}
A generalization of these estimates to
$S_{n,m}$ implies that in the confining phase there is the so
called monopole condensation and in the weak coupling region the
physical Hilbert space ${\cal H}$ decomposes into orthogonal sectors
labelled by total magnetic charge.

\section{Monopole condensation in the confining phase}
\label{condensation}
In this section we are going to prove monopole condensation in the strong
coupling regime: this relies on the fact that given arbitrary $\gamma
\in ]0,1[ $, there exists a $\beta_{\gamma}$ such that for $\beta \leq
\beta_{\gamma}$ holds
\begin{equation}
\left. S_{2} \, (x,q \, ; \, y,-q) \geq \gamma \ \ \ \ \ \forall \ x,y \in
\Lambda \ . \right.
\label{cond0}
\end{equation}
 In order to prove our statement we shall adopt the following strategy
\cite{KennKing}: first, we express the disorder field expectation
value in terms of logarithms of the partition functions
\begin{equation}
 \left.  S_{2} \, (x,q \, ; \, y,-q)  = \exp{\left[ \log Z(D) -
\log Z \right]} \right.
\label{cond6}
\end{equation}
and then we prove that $\left[ \log Z(D) - \log Z \right]$ is close to
 $0$ uniformely in $x$ and $y$.  The main tool we'll use is a cluster
 expansion for $\left[ \log Z(D) \right]$, which we obtain after
 rearranging $Z(D)$ as the partition function of a polymer gas.

\subsection{The polymer expansion}
The first step in our program is the polymer expansion for the
 (modified) partition function of our system (see
 e.g. \cite{Bryd,Seil}). Although it is a standard technique we now
 briefly present its application to the compact $U(1)$ system with
 disorder fields.  First note that $Z(D,\Omega)$ (\ref{mix2}) can be
 written summing over closed integer valued 2-forms $v$ as follows:
\begin{equation}
\left.  Z \, (D,\Omega) = N_{\Lambda} \, \sum_{v \, : \, dv = 0} \
 \prod_{ p \subset \Lambda} \tilde{I}_{\beta} \, (v_{p}) \ \ e^{i 2 \pi q
 (A_{v} ,D)} \ \ , \right.
\label{cond1}
\end{equation}
with
\begin{equation}
\left.  N_{\Lambda} = \left[ I_{\beta} (0) \right]^{N_{P}(\Lambda)} \
\ \ \ \ \ \ \ \ \ \ \ \ \ \tilde{I}_{\beta} \, (v_{p}) =
\frac{I_{\beta} (v_{p})}{I_{\beta} (0)}  \ \ . \right.
\label{cond1.5}
\end{equation}
In equation (\ref{cond1}) $A_{v}$ is the representative element of the
class defined by $d A_{v} = v$: it is simple to verify that each term
in the expansion does not depend on the choice of the representative
element $A_{v}$. Moreover we have extracted an overall factor
rescaling the modified Bessel functions by $I_{\beta}(0)$: the
advantage of this choice is that $\tilde{I}_{\beta}(0) = 1$ and in our
expansion around $v \equiv 0$ we must not carry over tedious
factors. \\ Let us now give some definitions: the support of a k-form
$v^{(k)}$ is the following set:
 $$ \mbox{supp} \, v^{(k)} = \left\{ x \, \in \, \Lambda : x \, \in \,
c_{k} \ \, \mbox{k-cell} \ \ \mbox{with} \ \ v^{(k)} (c_{k}) \neq 0
\right\} \ . $$ Let us note that following \cite{KennKing} we think of
supp$v^{(k)}$ as a set of points rather then a set of k-cells. In the
same way one can define in a natural way the support of a set of
k-cells. This permits us to extend to these sets the common definition of
connectedness: a set $X \subset \Lambda $ is connected if any two
sites in $X$ can be connected by a path of links whose endpoints all
lie in $X$.\\ Returning to equation (\ref{cond1}), in order to
recover the equivalent polymer system, we can rearrange it as
\begin{equation}
\left. Z(D) = \sum_{ v \,  : \, dv = 0 } \,   k(v)
\right. \ ,
\label{cond2}
\end{equation}
with (we miss the ininfluent factor $N_{\Lambda}$)
\begin{equation}
\left.  k(v)  =
 e^{2 \pi q i \, \left( A_{v}, D \right)} \, \prod_{p \subset \Lambda}
\, \tilde{I}_{\beta} \, (v_{p})  \right. \ .
\label{kappadivw}
\end{equation}
The main idea is to reexpress equations (\ref{cond2}) and
 (\ref{kappadivw}) in terms of closed 2-forms $v$ with connected
 supports, which will become the supports of the polymers. With this
 purpose let us recall that it is possible to write \cite{KennKing} $v
 = \sum_{i} v_{i}$ with the property that supp$\, v_{i}$ are the
 connected components of supp$\, v$. Moreover one can write $A_{v} =
 \sum_{i} A_{v_{i}}$ with $d A_{v_{i}} = v_{i}$ and these relations
 allow the following factorisation:
\begin{equation}
k(v) = \prod_{i} \, k(v_{i}) \ .
\label{cond3}
\end{equation}
Finally, observing that with the above definition of connected set
 the condition  $dv = 0$ implies  $dv_{i} = 0 \ \ \forall \,
i$, we can reorganise the sum in equation (\ref{cond2}) as follows:
\begin{eqnarray}
 \left. Z (D) = \sum_{n = 0}^{\infty} \, \frac{1}{n \, !} \sum_{X_{1}
...X_{n}} \, \prod_{i=1}^{n} \, K(X_{i},D) \right. & \ \ \ \ \ \ \ \ \
\ & \left. K(X,D) = \sum_{v: \, supp \, v = X} \, k(v) \right.  \ .
\label{cond4}
\end{eqnarray}
The sum is extended to all finite connected subsets $X_{i} \subset
 \Lambda$, with the condition that $X_{i} \, \cup \, X_{j}$ is
 disconnected if $i \neq j$. Thus we see that expression (\ref{cond4})
 has the form of a hard core interaction \cite{Bryd} between polymers
 $X_{i}$ of connected support. This allows us to use the techniques
 developed to deal with these systems in order to give the wanted
 bound on disorder field expectations. We point out that an expansion
 of the form given in (\ref{cond4}) is common to many statistical (or
 quantum mechanical in the lattice formulation) systems: the polymer
 activities $K(X_{i},D)$ are the link to the original problem,
 depending on the form of the starting action. In particular the same
 expansion is possible for the $U(1)$ gauge model with Villain action
 \cite{Vill} and gives monopole condensation in strong coupling
 \cite{FM2,FM1}. The only difference from Wilson model is that the
 dual representation is gaussian and this simplifies the handling of
 polymer activities.

\subsection{Bounds on polymer activities}
It is a well known result that the main properties (mathematical and
 physical) of the polymer system can be taken in strict correspondence
 with the general behavior of the activities, which in turn depends on
 parameters such as the temperature or the coupling constant. In this
 subsection we give a bound on $|K(X,D)|$ which is known \cite{Bryd}
 to be a sufficient condition for the convergence of the Mayer
 expansion for $\log Z(D)$; moreover it will be of great importance in
 the estimate of $\left[ \log Z(D) - \log Z \right]$.  We want to show
 that $\forall \ M > 0 \ \ \exists \ \beta_{M} \ $ such that for $
 \beta < \beta_{M}$
\begin{equation}
\mbox{(i)} \ \ \ |K(X,D)| \leq e^{- M \, |X|} \ \ \ \ \ \ \ \ \ \ \
\mbox{(ii)} \ \ \ | \frac{\partial}{\partial D_{b}} K(X,D) | \leq e^{-
M \, |X|} \ .
\label{cond7}
\end{equation}
Here $|X|$ stands for the cardinality of the support of the $X$ polymer.
The basic tool used in the proof of equation (\ref{cond7}) is the
following upper bound on the modified (and rescaled) Bessel functions,
\begin{equation}
\tilde{I}_{n} \, (\beta) \leq e^{|n|} \, \left( \frac{
 \beta}{2}\right)^{|n|} \ \ \ \ \ \ \ \ \ \ 0 \leq \frac{\beta}{2}
 \leq 1 \ ,
\label{cond8}
\end{equation}
which easily follows from the power expansion \cite{Grad}:
$$ I_{n} \, (\beta) = \left( \frac{\beta}{2}\right)^{n} \,
 \sum_{k=0}^{\infty} \, \frac{1}{\Gamma (k+1) \, \Gamma (n + k + 1)}  \,
 \left( \frac{\beta}{2}\right)^{2k} \ .  $$ The bound (\ref{cond8})
 tells us that for small $\beta$ the series whose $n$-th term is given
 by $I_{\beta} (n)$ is convergent.  Now let us sketch the argument
 that leads to (\ref{cond7}). In what follows we assume that supp$\,v
 = X$: from equations (\ref{kappadivw}) and (\ref{cond4}) one has
\begin{equation}
\left.   |K(X,D)| \leq \sum_{v \,: \, dv=0} \, \prod_{p \subset X}
\, | \tilde{I}_{\beta} \, (v_{p}) | \leq  \sum_{ all \ v} \,
\prod_{p \subset X} \, | \tilde{I}_{\beta} \, (v_{p}) | \right. \ .
\label{cond9}
\end{equation}
We have added the contributions due to all integer valued 2-forms on
$X$ because we have in mind to exploit the convergence property of
$\sum I_{\beta}(n)$. In order to sum over all integer 2-forms with
support on $X$, we first consider each set $Y_{i}$ of plaquette such that
 supp$Y_{i} = X$ and sum over 2-forms $v$ satisfying $v_{p} \neq 0 \ \
\forall \,  p \in Y_{i}$. Then we collect the contributions due to all the
sets  $Y_{i}$. Formally we have:
\begin{equation}
\left.  |K(X,D)| \leq \sum_{Y_{i}: supp \, Y_{i} = X} \
 \sum_{v: v_{p} \neq 0} \, \prod_{p \in Y_{i}} \, \left[
 \tilde{I}_{\beta} \, (v_{p}) \right]  \right. \ .
\label{cond9.5}
\end{equation}
Now let us focus on the generic term with fixed $Y_{j}$: exchanging
the sum with the product and using the parity property $I_{n} \,
(\beta) = I_{-n}\, (\beta) $, we obtain for the polymer activity
\begin{equation}
\left.   |K(X,D)| \leq   \sum_{Y_{i}: supp \, Y_{i} = X} \
\prod_{p \in Y_{i}} \left[ 2 \, \sum_{n>0} \,
 \tilde{I}_{\beta} \, (n) \right]  \ . \right.
\label{cond10}
\end{equation}
Thus to each plaquette in $Y_{i}$ is associated a factor that we extimate
 replacing $I_{\beta} (n)$ by the upper bound (\ref{cond8}) and
 summing the resulting series. Noticing that this is a geometric
 series missing the first term we are left with
\begin{equation}
\left. \prod_{p \in Y_{j}} \, \frac{e \beta}{1- \frac{e \beta}{2}}
 \leq e^{- N_{p}(Y_{j}) \, \log (\frac{1}{2 e \beta})}  \ \ \ \ \ \ \ \ \ \
 \mbox{for} \ \ \ \beta < \frac{1}{e} \ , \right.
\label{cond11}
\end{equation}
where $N_{p} (Y_{j})$ is the number of plaquettes in $Y_{j}$.
From the fact that supp$Y_{j} = X$ follows the relation $N_{P}(Y_{j}) \geq
 \frac{1}{4} |X|$. Moreover one can bound the number of $Y_{j}$ with
support on $X$ by $e^{k \, |X|}$  and obtains
\begin{equation}
\left.   |K(X,D)| \leq  e^{- (A_{\beta} - k) \, |X|} \ \ \ \ \ \ \ \ \ \
A_{\beta} = \frac{1}{4} \, \log (\frac{1}{2 e \beta }) \right. \ .
\label{cond12}
\end{equation}
Thus we see that part (i) of (\ref{cond7}) is satisfied with
$M_{\beta} = A_{\beta} - k$. As far as part (ii) of (\ref{cond7})
is concerned, it can be obtained showing that there is a function
$G(\beta)$ such that holds
\begin{equation}
\left.  |\frac{\partial}{\partial D_{b}} K(X,D)| \leq G(\beta) |X|^{3} \, e^{-
(\frac{|X|}{4} - 1) M_{\beta} } \equiv  F(|X|) \ ; \right.
\label{cond13}
\end{equation}
then one finds a constant $M'_{\beta}$ such that
\begin{equation}
F(|X|) \leq e^{-M' \, |X|} \ .
\label{cond14}
\end{equation}
For a proof of equation (\ref{cond13}) see appendix \ref{apa}. The proof of
(\ref{cond14}) can then be obtained by means of elementary analysis.

\subsection{Bound on $S_{2} \, ( x,q \, ; \, y,-q)$}
For sake of completeness let us  now sketch the analysis given in
\cite{KennKing} to bound $\left[ \log Z(D) - \log Z \right]$.
Let us start from the definition of the Mayer expansion:
\begin{equation} \left.
\log Z(D) = \sum_{n =0}^{\infty} \frac{1}{n \, !} \, \sum_{X_{1}...X_{n}} \,
\psi_{c}(X_{1}...X_{n}) \, \prod_{i = 1}^{n} \, K(X_{i},D) \ . \right.
\label{bub1}
\end{equation}
Here $\psi_{c}(X_{1}...X_{n})$ is the connected part of the hard core
interaction of the polymer system and it is nonvanishing only if
$\cup_{i} \, X_{i}$ is a connected set. The bound on the polymer
activities allows the convergence of the Mayer expansion \cite{Bryd}
and this implies, for example, that correlation functions defined by
differentiation of $\log Z(D)$ with respect to $D_{b}$ share the
cluster property. \\
Following the line of \cite{KennKing} we now define $H(s) = \log
Z(sD)$ and observe that $H(s) = H(-s)$ because the measure on equivalence
classes $[A]$ is even. Hence we see that
\begin{equation} \left. \log Z(D) - \log Z(0) = \int_{0}^{1} \, ds
\,(1-s) \, H''(s) \ ,\right.  \label{bub2} \end{equation} where
\begin{equation} H''(s) = \sum_{b,b'} \, D(b) D(b') \, m(b,b')
\label{bub3} \end{equation}
\begin{equation}
\left.  m(b,b') = \sum_{n = 0}^{\infty} \frac{1}{n \, !} \, \sum_{i,j
 = 1}^{n} \, \sum_{ b \subset B(X_{i}), b' \subset B(X_{j})} \,
 \psi_{c} (X_{1}...X_{n}) \, \times \left[ \prod_{k \neq i,j} \,
 K(X_{k},sD) \right] \frac{\partial}{\partial D_{b}} K (X_{i},sD) \,
\frac{\partial}{\partial D_{b'}} K (X_{j},sD) \ .  \right.
\label{bub4}
\end{equation}
In (\ref{bub4}) $B(X)$ denotes the smallest rectangular parallelepiped
 in $\Lambda$ which contains $X$. Moreover, by the exponential bounds
on activities and derivatives given in equation (\ref{cond7}) follows
that
\begin{equation}
\left. \left[ \prod_{k \neq i,j} \, K(X_{k},sD) \right]
 \frac{\partial}{\partial D_{b}} K(X_{i},sD) \,
 \frac{\partial}{\partial D_{b'}} K(X_{j},sD) \leq \exp{ \left[ - M \,
 \sum_{k} |X_{k}| \right] } \ . \right.
\label{bub4.5}
\end{equation}
Since  $\psi_{c} (X_{1}...X_{n}) \neq 0$ only if  $\cup_{i} \, X_{i}$
is a connected set, we have that  $\sum_{k} |X_{k}| \geq d(b,b')$
($d(b,b')$ is the distance between the two links) and so
\begin{equation}
 \left.  |m(b,b')| \leq \exp{\left[ - \frac{1}{2} M d(b,b') \right]}
  \times \, \sum_{n = 0}^{\infty} \frac{1}{n \, !} \, \sum_{i,j = 1}^{n}
  \, \sum_{ b \subset B(X_{i}), b' \subset B(X_{j})} \, \psi_{c}
  (X_{1}...X_{n}) \, \exp{\left[ - \frac{1}{2} M \, \sum_{i} \,
  |X_{i}| \right] } \right. .
\label{bub5}
\end{equation}
Now, using the techniques of \cite{Bryd}, one can work out from
 (\ref{bub5}) the inequality
\begin{equation}
 \left.  |m(b,b')| \leq \exp{\left[ - \frac{1}{2} M d(b,b') \right]}
\times \, \sum_{n} \, n \, \delta^{n} \sum_{ X \supset b }
|X| \, \exp{\left[ - \frac{1}{2} M \, |X| \right]} \ ,   \right.
\label{bub6}
\end{equation}
where $\delta$ is a constant little as $\beta$ decreases. For $M$
large enough the sum over $n$ appearing in (\ref{bub6}) converges to a
constant $\delta'$ which again is little as $\beta$ decreases.
Moreover reading the sum in (\ref{bub3}) as scalar product between
 $D$ and  $m D$, we have the following bound on $|H''(s)|$:
 \begin{equation}
|H''(s)| = |(D,m \, D)| \leq
\|D\|_{2} \, \|m \, D\|_{2} \leq \|D \|^{2}_{2} \, \|m\|_{2} \ .
\label{bub8}
\end{equation}
From properties on matrix norms we have that
$\|m\|_{2} \leq sup_{b'} \, (\sum_{b} |m (b,b')| )$: using the previous
result we obtain
\begin{equation}
\left. |H''(s)| \leq \delta'
\, \|D\|^{2}_{2} \left[ \sum_{b} e^{- \frac{1}{2} M \, d(b,b')}
\right] < \rho (\beta) \ , \right.
\label{bub9}
\end{equation}
with $\rho (\beta) \rightarrow 0$ as $\beta \rightarrow 0$. It is
 important to point out that the argument works because $\|D\|_{2}$ is
 bounded uniformely in $x$ and $y$.  Finally, using (\ref{bub2}) and
 (\ref{bub3}) we conclude
\begin{equation}
\left. S_{2} (x,q \, ; \, y -q) > \exp{ \left[ - \frac{1}{2} \, \rho
(\beta) \right] } \right. \ .
\label{bub10}
\end{equation}
This relation implies in particular that $S_{1} (x,q)$, defined by
 the limiting procedure in which $y \rightarrow \infty$, is nonvanishing.
In the language of field theory we can say that the field
 describing magnetic monopoles aquires a nonvanishing vacuum expectation
value. This implies the spontaneous breakdown of the topological
 symmetry associated to the magnetic charge conservation and signals
 confinement of electric charge \cite{thooft}.

\section{Monopole sectors in the QED phase}
\label{sectors}
In this section we prove the relation (\ref{cluster}) for the soliton
 two point function, showing that for $\beta$ large enough and
$|x-y| \rightarrow \infty$ it is possible to find a positive constant
 $m(\beta)$ such that
\begin{equation}
\left. S_{2} \, (x,q \, ; \, y,-q) \leq e^{ - m(\beta)
q |x-y|} \ . \right.
\label{ww1}
\end{equation}
In order to proof our statement, we use a slight modification of the
 expansion given in \cite{FS1,FS2} and \cite{tesi}: we reexpress the
 partition function as a gas of monopole loops, to which apply a
 renormalization transformation. Estimates on the renormalized loop
 activities enable us to extract the relevant contribution to $S_{2}
 (x,q \, ; \, y,-q)$.

\subsection{Expansion of $Z(D,\Omega)$ in interacting  monopole loops}
From equation (\ref{mix2}), defining the modified partition function
in dual representation, it is natural to introduce a measure on
equivalence classes $[A]$ of integer valued 1-forms given by
\begin{equation}
\left. d\mu \, (A) = \frac{1}{Z} \, \prod_{p \subset \Lambda} \,
I_{\beta}\, (dA_{p}) \ \ \ \ \ \ \ \ \ \ \ \ \ \int_{[A]} d\mu \, (A)
= 1 \ .  \right.
\label{ww3}
\end{equation}
The main idea \cite{FS1} on which is based our construction is the
following: we want to introduce a measure $d \mu_{I_{\beta}} \, (A) $
on $R^{n}$ (n is the number of links in $\Lambda$), which reproduces
(\ref{ww3}) once we constrain the real variables $A_{b}$ to integer
values and pick them on a gauge slice. Such a measure should enable us
to make suitable extimates in the weak coupling region. We can fulfill
the first constraint inserting a sum of $\delta$-functions for each
link variable $A_{b}$, which by Poisson summation gives us the
monopole currents; formally we have
\begin{equation}
\left.  d\mu \, (A) = \frac{1}{ Z} \,  \prod_{b \subset \Lambda}
\, \left[ \sum_{q \in Z} \, e^{ i 2 \pi \, q \, A_{b}} \right] \,
 d \mu_{I_{\beta}} \, (A) \ .
\right.
\label{ww4}
\end{equation}
% $$ \sum_{\begin{array}{c}pippo\\pappo \end{array} }     $$
%Moreover the gauge choice $\delta \, A = 0$ should be implemented by the
%following constraint, which selects monopole loops in (\ref{ww4}):
Moreover, since we are going to compute expectations values of gauge
invariant observables, the only contributions come from conserved
currents; hence we can impose
\begin{equation}
\left.   \int d \mu_{I_{\beta}}  (A) \, e^{i \, (J,A)} = 0 \ \ \ \ \ \ \ \ \
\ \ \ \ \ \ \ \ \ \ \ \ \mbox{for} \ \ \ \delta \, J \neq 0 \ . \right.
\label{ww6}
\end{equation}
Actually it is possible to construct a measure with the required properties,
taking  the limit $\Lambda \rightarrow Z^{4}$ of
\begin{equation}
\left. d \mu_{I_{\beta}} \, (A) = \frac{1}{N_{\Lambda}} \,
\prod_{p \subset \Lambda} \, I_{\beta} \, (dA) \, \prod_{b \in \Lambda}
\, d A_{b}  \ \ \ \ \ \ \ \ \ \ \ \ A_{b}   \in  R \ \ . \right.
\label{ww7}
\end{equation}
 $I_{\beta} \, (\phi)$ is a suitable interpolation of the modofied
 Bessel functions, which has been constructed in \cite{FS1} for large
 values of $\beta$, and has the properties listed below:
\begin{enumerate}
\item $$ I_{\beta} \, (\phi) = \frac{1}{2 \pi} \, \int_{0}^{2 \pi} \,
d \theta \, e^{ \beta \cos \theta} \, e^{i \phi \theta} \ \ \ \ \
\ \ \ \ \ \ \ \mbox{for integer } \ \phi  \ ; $$
\item $I_{\beta} \, (\phi)$ is an  {\em even}, {\em positive} and
{\em integrable} function on  $R$;
\item  $I_{\beta} \, (\phi)$ is  {\em analitic} on the strip $|{\cal I}$m$ \,
\phi | \leq \frac{\beta}{2}$. Moreover it is possible to find a constant $c$
such that
$$ \left. | \frac{ I_{\beta} \, (\phi+ ia)}{ I_{\beta} \, (\phi)} |
\leq \exp{\left[\frac{g(a)}{\beta} \right]} \ \ \ \ \ \ \ \ \ \ \ \ \
\mbox{with} \ \ \ \ 0 \leq g(a) \leq c \, e^{2 \pi |a|} \ \ \ \ \ \ \
\mbox{for} \ \ 1 \leq |a| \leq \frac{\beta}{2} \ . \right. $$
\end{enumerate}
With these new tools the soliton two point function can be written
\begin{equation}
\left. S_{2} \, (x,q \, ; \, y,-q) = \int d \, \mu_{I_{\beta}}  (A)
e^{i 2 \pi q (A, (D - \Omega))} \, \prod_{b \in \Lambda} \left[ \,
\sum_{n_{b}} e^{i 2 \pi \, n_{b} A_{b}} \right] \right. \ .
\label{ww8}
\end{equation}
Henceforth, for notational convenience, the factor $2 \pi q$ will be
absorbed in $D$ and $\Omega$. Our first objective is a suitable
rearrangement of the product appearing in (\ref{ww8}), which, by
the constraint (\ref{ww6}), gives the usual coupling of $A$ with
external current loops.\\ Let us start defining a current density $\rho$
as a 1-form on $\Lambda$ with values in $2 \pi Z$.  An 1-ensemble
${\cal E}$ is a set of current densities $\{ \rho \}$ whose supports
are disjoint and such that dist$(\rho, \rho') \geq 2^{\frac{1}{2}} \ \
\forall \rho , \rho '  \in \, {\cal E} $. It is useful to collect
the above currents in 1-ensembles using the following property
\cite{FS2}:
\begin{equation}
\left.  \prod_{b \in \Lambda} \left[ \, \sum_{n_{b}} e^{i 2 \pi \,
n_{b} A_{b}} \right] = \sum_{\sigma} \, d_{\sigma} \,
\prod_{\rho \in {\cal E}_{\sigma}} \, \left[ 1 + K(\rho) \, \cos (A,\rho)
\right]  \right.
\label{ww9}
\end{equation}
The index $\sigma$ runs over a finite set, each ${\cal E}_{\sigma}$ is
an 1-ensemble and $ d_{\sigma} > 0 $. Moreover, the bare loop
activities satisfy $ 0 < \, K(\rho) < \, e^{k \|\rho\|_{1}}$, where
$\|\rho \|_{1}$ is the norm given by $\|\rho\|_{1} = \sum_{b \in supp
\, \rho} \, |\rho_{b}|$. The next step is to consider the string
$\Omega$ as a current density and construct suitable 1-ensembles
containing open-ended currents which are obtained by grouping the
currents ``touching'' $\Omega$, in a sense which we'll specify below.
We choose  $\Omega$ (by hyper-gauge invariance) such that $\mbox{supp} \,
\Omega \cap \, \mbox{supp} \, D = \oslash$:
This means that we are taking $\Omega$ with support in the region bounded by
the hyperplanes $z^{0} = x^{0}$ and $z^{0} = y^{0}$. The disorder
field expectation can be written as follows \cite{FS1,FS2,tesi}:
\begin{equation}
\left. Z(\Omega,D) = \int \, d \mu_{I_{\beta}} \, (A) \, \sum_{\tau}
 \, c_{\tau} \, \left[ \cos (A, D - \Omega) + K(\rho^{\tau}) \, \cos
 (A, D - \Omega + \rho^{\tau} ) \right] \times \prod_{\rho \in {\cal
 E}_{\tau}} \, \left[ 1 + K(\rho) \, \cos (A, \rho) \right] \
 . \right.
\label{ww10}
\end{equation}
Now, some comments on equation (\ref{ww10}): the currents
$\rho^{\tau}$ are divergenceless and their support has non vanishing
intersection with $ N = \left\{ b \in \Lambda \  : \  \mbox{dist}
\,(b, \mbox{supp} \, (D - \Omega)) \leq 1 \right\}$: $\rho^{\tau}$ are
the currents touching $\Omega$.  $c_{\tau}$ are positive constants and
$K (\rho^{\tau})$ satisfies $ 0 < \, K(\rho^{\tau}) < \, e^{k
\|\rho^{\tau} \|_{1}}$.  Moreover $ {\cal E}_{\tau}$ is an 1-ensemble
of divergenceless currents having vanishing intersection with $N$. We
must point out that ${\cal E}_{\tau} \cup \left\{ \rho^{\tau} - \Omega
\right\}$ is still an 1-ensemble: we have divided the closed currents
from the open-ended $\rho^{\Omega_{\tau}} = \rho^{\tau} - \Omega$, which
play a peculiar role in the proof of clustering of the soliton
correlation function, as we shall see below.

\subsection{Renormalization transformation}
Now we make on (\ref{ww10}) a transformation that renormalizes the
activities of currents and allows us to give the wanted bound on
$S_{2} (x,q \, ; \, y,-q)$. The transformation consists in the
explicit integration of the factors containing the currents $\rho$ on
a suitable subset of supp$\, \rho$, which we call ${\cal B}_{\rho}$,
characterised by the fact that two different links contained in it
belong to different plaquettes and
\begin{equation}
\left. \sum_{b \in {\cal B}_{\rho}} \, |\rho_{b}| \geq \tilde{c} \,
\|\rho \|_{1}  \ . \right.
\label{wrin0}
\end{equation}
In dimension four the geometric constant $\tilde{c}$  can be fixed to be
$\frac{1}{18}$. Such a renormalization is based on the application of the
following property. Let us consider a function $G(A)$ which does not
depend on $A_{b}$ for some link $b \in \Lambda$: for arbitrary real $a$ such
 that  $ \, |a| \leq \frac{\beta}{2}$ then we have
\begin{equation}
\left. \int \, e^{i q A_{b}} \, G(A) \, d \mu_{I_{\beta}} \, (A) =
  e^{- \tilde{E}_{\beta}(a,q)} \, \int \, e^{ i q A_{b}} \, \left[
  \prod_{p : b \in \partial p} \, i_{\beta} \, (\epsilon (b,p) a , dA
  ) \right] \, G(A) \, d \mu_{I_{\beta}} \, (A) \ , \right.
\label{wrin1}
\end{equation}
where
\begin{eqnarray}
\left. \tilde{E}_{\beta} (a,q) = q a - \frac{n_{b}}{\beta} \, g(a)
\right. & \ \ \ \ \mbox{and} \ \ \ \ & \left. i_{\beta} \, (a, \phi) =
\frac{I_{\beta} \, (\phi + ia)}{I_{\beta} \, (\phi)} \, e^{-
\frac{g(a)}{\beta}} \ . \right.
\label{wrin1.5}
\end{eqnarray}
In equation (\ref{wrin1.5}) $ n_{b}$ is the number of plaquettes
containing $b$ and $\epsilon (b,p)$ is a factor $\pm 1$ which gives the
orientation of $b$ in $\partial p$. The proof of this property is
immediate if we make a complex translation on the variable $A$ ( $A
\rightarrow A + ia$) and use the properties of the function $I_{\beta}
(\phi)$.\\ Now let us focus on the generic term in the sum over the
index $\tau$ appearing in (\ref{ww10}). First, we choose ${\cal
B}_{\rho} \subset \mbox{supp} \, \rho $ for every $\rho \in {\cal
E}_{\tau}$ and ${\cal B}_{\rho^{\tau}} \subset \mbox{supp} \,
\rho^{\Omega_{\tau}}$. By the defining property of these subsets and
the fact that we are dealing with 1-ensembles, all links selected in
this way belong to different plaquettes.  After exponentiating the
cosines in (\ref{ww10}) and decomposing the resulting products, this
observation allows us to apply the relation (\ref{wrin1}) to the
resulting terms for all links in ${\cal B}_{\rho}$, ${\cal
B}_{\rho^{\tau}}$ and $\mbox{supp} \, \Omega$. These integrations
produce the transformation of $\cos (A,\rho)$ into a function
$c_{\beta} (A,\rho)$ and renormalize the activities $K(\rho)$ as
described by the following relations:
\begin{equation}
 \left.  Z(\Omega,D) = \int \, d \mu_{I_{\beta}} \, (A) \, \sum_{\tau}
 \, c_{\tau} \, \left[ z(\beta) c_{\beta} (A, D - \Omega) +
 z(\beta,\rho^{\tau}) \, c_{\beta} (A, D + \rho^{\Omega_{\tau}})
 \right] \times \prod_{\rho \in {\cal E}_{\tau}} \, \left[ 1 +
 z(\beta,\rho) \, c_{\beta} (A,\rho) \right] \ .\right.
\label{wrin2}
\end{equation}
The renormalized activities are given by
\begin{eqnarray}
z(\beta) & = & \prod_{b \in \ supp \, \Omega} \,  e^{- \tilde{E}_{\beta} \,
(a,\Omega_{b}) }  \ ; \nonumber \\ z(\beta,\rho^{\tau}) & = &
K(\rho^{\tau}) \, \prod_{b \in {\cal B}_{\rho}^{\tau}}  \, e^{-
 \tilde{E}_{\beta} \, \left( a , (D  + \rho^{\Omega_{\tau}})_{b} \right) }
 \ ; \nonumber \\
 z(\beta,\rho) & = & K(\rho) \, \prod_{b \in {\cal B}_{\rho}} \,
  e^{- \tilde{E}_{\beta} \, (a,\rho_{b}) } \ .
\label{wrin2.5}
\end{eqnarray}
The renormalized version of the cosine is
$ c_{\beta} \, (A,\rho) = {\cal R}e \left[ e_{\beta} \, (A,\rho) \right] $
where
\begin{equation}
\left. e_{\beta} \, (A,\rho) = e^{i (A,\rho)} \left[ \prod_{p \in
T({\cal B}_{\rho})} \, i_{\beta} \, \left( \epsilon a ,dA(p) \right)
\right] \right.  \ \ \ \ \ \ \ \ \ \ \ \mbox{and} \ \ \ \ \ \ \ \ \ \
\ \left. T({\cal B}_{\rho}) = \left\{ p \in \Lambda \ \, : \  b \in
 \partial p \ \ \  b \in  {\cal B}_{\rho} \right\} \ . \right.
\label{wrin3}
\end{equation}
From the relations given above it is clear that $ | c_{\beta} \, (A ,
 \rho) | \leq 1$.

\subsection{Estimates on renormalized activities and cluster property}
In order to extract a bound that assures the cluster property we must
now choose a suitable value of the parameter $a$ appearing in
$\tilde{E}_{\beta}$.  By property $3.$ of the function $I_{\beta} \,
(\phi)$ follows that
\begin{equation}
 \left. e^{ - \tilde{E}_{\beta} \, (a,q)} \leq e^{ - E_{\beta} \,
 (a,q)} \ \ \ \ \ \ \mbox{with} \ \ \ \ \ \ \ E_{\beta} \, (a,q) = q a
 - \frac{n_{b}}{\beta} \, c \, e^{2 \pi |a|} \ .  \right.
\label{watt1}
\end{equation}
In order to give an upper bound on activities as strong as possible, we take
 the value af $a$ maximizing $E_{\beta} (a,q)$ in the domain $|a| \leq
 \frac{\beta}{2}$ and we denote it by $a_{m}$. For a {\em fixed
 value} of $\beta$ it turns out that $a_{m}$ depends on the value of
 the parameter $q$, which stands here for the value of the generic
 current on a link in ${\cal B}_{\rho}$ ($q \rightarrow \rho_{b}$ in
 equations (\ref{wrin2.5})). We find
\begin{eqnarray}
\left. a_{m}^{(1)} (q) = \epsilon (q) \, \frac{1}{2 \pi} \log \left(
 \frac{ \beta \, |q|}{2 \pi n_{b} c} \right)  \right.  \ \
 \mbox{if} \ |q| \leq \tilde{q}_{\beta} ;  &   \ \ \ \ \ \ \ \ \ \ \ &
\left.  a_{m}^{(2)} (q) =  \epsilon (q) \, \frac{\beta}{2} \ \ \mbox{if} \
 |q| \geq  \tilde{q}_{\beta},   \right.
\label{watt2}
\end{eqnarray}
and correspondingly
\begin{eqnarray}
\left.  E_{\beta}^{(1)} \, (q) \equiv E_{\beta} \, (a_{m}^{(1)} ,q) =
 \frac{|q|}{2 \pi} \, \left[ \log \left( \frac{ \beta \, |q|}{2 \pi
 n_{b} c} \right) \, - 1 \right] \, ; \right.  & \  \ \ \ \ \ \ \ \ \
 & \left. E_{\beta}^{(2)} \, (q) \equiv E_{\beta} \, (a_{m}^{(2)} ,q)
 = |q| \, \frac{\beta}{2} \left( 1 - \frac{2 n_{b} c}{|q| \beta^{2}}
 \, e^{\pi \beta} \right) \ . \right.
\label{watt2.2}
\end{eqnarray}
The discriminant value $\tilde{q}$ is defined by $ a_{m}^{(1)} (\tilde{q}) =
\frac{\beta}{2}$.  The important feature of these  current
self-energies exctracted by renormalization is that for $\beta$ large
enough both $E_{\beta}^{(1)} \, (q)$ and  $E_{\beta}^{(2)} \, (q)$ are
 positive. \\
With the above choice for the parameter $a$ the renormalized activity of
 the generic current $\rho$ satisfies
\begin{equation}
\left. z( \beta, \rho) \leq K(\rho) \, \prod_{b \in {\cal B}_{\rho}} \,
e^{ - |\rho_{b}| \, A(\beta, \rho_{b})} \, ;  \ \ \ \ \ \ \ \
A(\beta,\rho_{b}) = \cases{ \frac{1}{2 \pi} \, \left[ \log \left(
 \frac{ \beta \, |\rho_{b}| }{2 \pi n_{b} c} \right) \, - 1 \right] &for
 $|\rho_{b}| \leq \tilde{q}$ ,\cr &\cr \frac{\beta}{2} \left( 1 -
 \frac{2 n_{b} c}{|\rho_{b}| \beta^{2}} \, e^{\pi \beta} \right) &for
 $|\rho_{b}| \geq \tilde{q}$ } \right.
\label{watt3}
\end{equation}
In both cases $A(\beta,\rho_{b})$ is bounded from below by a function
which does not depend on $\rho_{b}$:
\begin{eqnarray}
A(\beta,\rho_{b}) \geq  \cases{ \frac{1}{2 \pi} \, \left[ \log \left(
 \frac{ \beta }{12 \pi  c} \right) \, - 1 \right] = A^{(1)} \, (\beta) &
 \cr &\cr \frac{\beta}{2} \left( 1 -
 \frac{1}{\pi  \beta} \right) = A^{(2)} \, (\beta) & } \ .
\label{watt3.5}
\end{eqnarray}
The first bound is obtained using the inequalities $|\rho_{b}| >1$
 and $n_{b} \leq 6$; the second one is obtained replacing $|\rho_{b}|$
 with $\tilde{q}$.
We must piont out that both $ A^{(1)} \, (\beta)$ and $ A^{(2)} \, (\beta)$
 are positive functions increasing with $\beta$. If now we  define
 $A \, (\beta) =   \mbox{min} \{  A^{(1)} \, (\beta),
 A^{(2)} \, (\beta) \}$, using the properties of ${\cal B}_{\rho}$
 and $K(\rho)$ we can write
\begin{equation}
\left. z( \beta, \rho) \leq K(\rho) \, e^{ - \tilde{c} \, A(\beta) \,
\| \rho \|_{1} } \leq e^{ - \left( \tilde{c} \, A(\beta) - k \right)
\, \| \rho \|_{1} } \ . \right.
\label{watt4}
\end{equation}
In particular one can bound the renormalized activity of the string as
follows:
\begin{equation}
\left. z( \beta ) \leq  e^{ - \tilde{c} \,  A(\beta) \, \| \Omega \|_{1} }
 \leq  e^{ -  \tilde{c} \, A(\beta) \, q  |x-y|}  \ . \right.
\label{watt5}
\end{equation}
More delicate is the estimate of $z(\beta,\rho^{\tau})$, because of
 the presence of the Coulomb field $D$. In fact one has ${\cal
 B}_{\rho^{\tau}} \subset \mbox{supp} (\rho^{\Omega_{\tau}})$ but the
 generalized current density is $\left( \rho^{\Omega_{\tau}} + D \right)$.
In general for these activities holds
\begin{equation}
\left. z( \beta, \rho^{\tau} ) \leq  C(\beta) \, e^{ -  \tilde{A}(\beta)
 \,  q |x - y| } \ . \right.
\label{watt6}
\end{equation}
The starting point to prove (\ref{watt6}) is the relation
\begin{equation}
 \left. z(\beta,\rho^{\tau}) \leq K(\rho^{\tau}) \, \prod_{b \in {\cal
B}_{\rho^{\tau}}} \, e^{- A(\beta) \, | \rho^{\Omega_{\tau}}_{b} +
D_{b} | } \ \ . \right.
\label{watt5.5}
\end{equation}
First one can easily see that in the case in which $\mbox{supp} \,
 \rho^{\tau} \cap \, \mbox{supp} \, D = \oslash$ one can deal with the
 current $\rho^{\Omega_{\tau}} = \rho^{\tau} - \Omega$ as for the common
 $\rho$ (see equation (\ref{watt4})) and from the relation
 $\|\rho^{\Omega_{\tau}} \| \geq q \, |x-y|$ follows (\ref{watt6}) with
 $C(\beta) = 1$. \\ The case in which $\mbox{supp} \, \rho^{\tau} \cap
 \mbox{supp} \, D \neq \oslash$ can be worked out with the following
 trick: we distinguish the links $b \in {\cal B}_{\rho^{\tau}}$ such
 that $|D_{b}| > \frac{1}{2}$ from those such that $|D_{b}| <
 \frac{1}{2}$ ( in other words we decompose ${\cal B}_{\rho^{\tau}} =
 {\cal B}_{\rho^{\tau}}^{<} \cup {\cal B}_{\rho^{\tau}}^{>}$ )
 obtaining a factorization in equation (\ref{watt5.5}).  Noticing then
 that $\rho^{\tau}$ takes values in $2 \pi Z$, for $ b \in {\cal
 B}_{\rho^{\tau}}^{<} $ we have
\begin{equation}
\left. |\rho^{\Omega_{\tau}}_{b} + D_{b} | = |\rho^{\Omega_{\tau}}_{b}| \, |
\left( 1 + \frac{D_{b}}{\rho^{\Omega_{\tau}}_{b}} \right) | \geq
|\rho^{\Omega_{\tau}}_{b}| \, \frac{1}{2} \ , \right.
\label{watt7}
\end{equation}
by which we reduce the factor involving ${\cal B}_{\rho^{\tau}}^{<}$ to
 the standard form (\ref{watt3}). The term involving ${\cal
 B}_{\rho^{\tau}}^{>}$ will give us the constant $C(\beta)$. In fact
 after little simple algebra we obtain
$$ \left. z(\beta, \rho^{\tau}) \leq K(\rho^{\tau}) \, e^{- \frac{1}{2}
 \, \tilde{c} \, A(\beta) \, \| \rho^{\Omega_{\tau}} \|_{1} } \, \, G(D,
 \rho^{\Omega_{\tau}}) \ \ , \right. \ \ \ \ \ \ \ \ \mbox{with} \ \ \ \ \
 \ \ \ \left.  G(D, \rho^{\Omega_{\tau}}) = \prod_{ b \in {\cal
 B}_{\rho^{\tau}}^{>} } \, \exp{\left[ A(\beta) \, \left(
 \frac{|\rho^{\Omega_{\tau}}_{b}|}{2} - | \rho^{\Omega_{\tau}}_{b} + D_{b}|
 \right) \right] } \ . \right. $$
Finally it is possible to show that $G(D, \rho^{\Omega_{\tau}})$ is bounded
 by a constant $C(\beta)$ only dependent on $\beta$ and the number of
 links on which $|D| \geq \frac{1}{2}$ (actually this number is a
 function of $q$); this completes the proof of equation (\ref{watt6}).

 Now we have all elements to complete our proof. By the property that
$|z c_{\beta} | \leq |z|$, starting from (\ref{wrin2}) we can write
\begin{equation}
\left. | S_{2} \, (x,q \, ; \, y,-q) | \leq \mbox{sup}_{\tau} \,
\left\{ z(\beta) + z(\beta, \rho^{\tau}) \right\} \times
\, \int d \mu_{I_{\beta}} \, (A) \, \sum_{\tau} \, c_{\tau} \,
\prod_{\rho \in {\cal E}_{\tau}} \, \left[ 1 + z(\beta,\rho) \,
c_{\beta} \, (A,\rho) \right] \ \ \ . \right.
\label{wconcl1}
\end{equation}
Let us call ${\cal R}$ the integral in (\ref{wconcl1}):
considering the explicit form of the normalization factor $Z$, it can be
represented as
\begin{equation}
\left. {\cal R} = \frac{ \sum_{\tau} \, A_{\tau}}{ \sum_{\tau} \, B_{\tau}}
 \ , \right.
\label{wconcl2}
\end{equation}
with
\begin{equation}
\left. A_{\tau} = c_{\tau} \, \int d \mu_{I_{\beta}} \, (A) \,
 \prod_{\rho \in {\cal E}_{\tau}} \, \left[ 1 +
 z(\beta,\rho) \, c_{\beta} \, (A,\rho) \right]  \, ; \right.
\label{wconcl3}
\end{equation}
\begin{equation}
\left. B_{\tau} = c_{\tau} \, \int d \mu_{I_{\beta}} \, (A) \,
 \left[ 1 + z(\beta,\rho^{\tau}) \, c_{\beta} \, (\rho^{\tau},A) \right] \,
 \prod_{\rho \in {\cal E}_{\tau}} \, \left[ 1 +
 z(\beta,\rho) \, c_{\beta} \, (A,\rho) \right]  \ . \right.
\label{wconcl4}
\end{equation}
It is simple to see that for $\beta$ and $|x - y|$ large enough we
have $ A_{\tau} \leq 2 \, B_{\tau}$.  In fact choosing suitable values
of $\beta$ and $|x - y|$ we can obtain that  $z(\beta,
\rho^{\tau}) \leq \frac{1}{2} $ and $z(\beta, \rho) \leq 1$; thus
 in order to evaluate $B_{\tau}$ we must integrate in the {\em positive}
measure
$$ d \mu_{I_{\beta}} \, (A) \, \prod_{\rho \in {\cal E}_{\tau}} \, \left[ 1
+ z(\beta,\rho) \, c_{\beta} \, (A,\rho) \right] $$
the function $\left[ 1 + z(\beta,\rho) \, c_{\beta} \, (A,\rho)
\right] \geq \frac{1}{2}$. This observation leads to the conclusion
that ${\cal R} \leq 2$. Finally using this result in (\ref{wconcl1})
 we get the relation
\begin{equation}
\left.  | S_{2} \, (x,q \, ; \, y,-q) | \leq 2 \ \mbox{sup}_{\tau} \,
 \left\{ z(\beta) + z(\beta, \rho^{\tau}) \right\} \ . \right.
\label{wconcl6}
\end{equation}
Cluster property of the soliton two point function follows then by
estimates of equation (\ref{watt5}) and (\ref{watt6}) on renormalized
activities.

\section{Concluding remarks}
\label{conclusion}
In conclusion we summarize the results: in the weak coupling region
 the Hilbert space of the reconstructed Lattice Quantum Field Theory
 splits into orthogonal sectors labelled by magnetic charge. Instead,
 in the strong coupling region the lattice solitons do condense in the
 vacuum sector: the symmetry associated to the topological
 conservation of magnetic charge is spontaneously broken, as signaled
 by the nonvanishing expectation value of the charged monopole
 operator. Indeed this is just the criterion for quark confinement
 proposed by 't Hooft \cite{thooft}.\\ Moreover our analytical results
 are in agreement with numerical calculations \cite{pisa2}: these show
 that the parameter regions in which $S_{1} > 0$ and $S_{1} = 0$
 coincide with the confining phase and QED phase, respectively, and
 allow to extract information about the behavior of the system at the
 transition. Our result suggests that the correlation functions of
 soliton (disorder) fileds are indeed a viable tool for the study,
 both numerical and analytical, of phase transitions in lattice models
 that exhibit monopole-like topological excitations (we point out that
 a slight generalization allows the extension to lattices with non
 trivial topology). Along this line there is the possibility to study
 analitically the disorder fields associated with vortices in the $3
 \, d$ $X \, Y$ model and, although more remote, a generalization to non
 abelian models.

\acknowledgements
We wish to thank Adriano Di Giacomo and Michele Mintchev for
 useful discussions.

\appendix
%\section{Bound on $|\frac{\partial}{\partial D_{b}} K(X,D)|$}
\section{Bound on activity's derivatives}
\label{apa}
In this appendix we sketch the argument which leads to equation
(\ref{cond13}).
The proof is based on the following inequality, given in \cite{KennKing}:
\begin{equation}
\left. \| A_{v} \| \leq \frac{c}{2 \pi}  \, (v,v)^{2} \leq c \,
N_{p}^{2} \, m_{v}^{4}  \ ,  \right.
\label{app1}
\end{equation}
where $N_{p}$ is the number of plaquettes in $X$ and
$$ \left. m_{v}  = \mbox{max}_{p \subset X} \, |v_{p}| \right.  \ . $$
The derivative with respect to $D_{b}$ modifies the expression of
$k(v)$ by the moltiplicative factor $ 2 \pi \, A_{v} (b)$ and thus we have:
\begin{equation}
\left. |\frac{\partial}{\partial D_{b}} K(X,D)| \leq c N_{p}^{2} \,
 \sum_{k > 0} \, k^{4} \, \sum_{[v]: m_{v} = k} \, \prod_{p \subset X}
 \, \tilde{I}_{\beta} \,(v_{p}) \right.
\label{app2}
\end{equation}
Now we can write
\begin{equation}
\left.  \sum_{[v]: m_{v} = k} \, \prod_{p \subset X}
 \, \tilde{I}_{\beta} \,(v_{p})  =  \sum_{Y_{i}: supp Y_{i} = X} \
  \sum_{[v]: m_{v} = k , v_{p} \neq 0} \, \prod_{p \in Y_{i}}
 \, \tilde{I}_{\beta} \,(v_{p}) \ . \right.
\label{app3}
\end{equation}
The 2-forms with $m_{v} = k$ may be obtained fixing the value of $v$
to $\pm k$ on a plaquette $p_{M}$ and summing over configurations of
integers $n \leq k$ in the remaining $(N_{P}(Y_{i}) -1)$ plaquette.
We extract a factor $\frac{\beta}{2}^{k}$ (coming from
$\tilde{I}_{\beta} (v_{p_{M}}= k)$) from each term with fixed $p_{M}$
and notice that the contribution due to all configurations is bounded by
$e^{- (N_{P}(Y_{i}) -1) \, A_{\beta}}$. The result is
\begin{equation}
\left.  |\frac{\partial}{\partial D_{b}} K(X,D)| \leq c N_{p}^{2} \,
 \sum_{k > 0} \, k^{4} \left( \frac{e \beta}{2} \right)^{k} \ \, e^{-
 (|X| - 1) \, A_{\beta}} \ \, \sum_{Y_{i}: supp Y_{i} = X} \,
 N_{P}(Y_{i}) \ . \right.
\label{app4}
\end{equation}
The series in $k$ can be estimated by the value of the integral of
$x^{4} e^{- B \, x}$ on $R_{+}$ ($B = \log (\frac{2}{e \beta})$), and
this gives $G(\beta)$ apart from factors. Moreover the sum over $Y_{i}$ is
bounded by $2 |X| e^{k|X|}$ in such a way that we can write equation
(\ref{cond13}):
\begin{equation}
\left.  |\frac{\partial}{\partial D_{b}} K(X,D)| \leq G(\beta) |X|^{3}
\, e^{- (|X| - 1) M_{\beta} } \equiv F(|X|) \ . \right.
\label{app5}
\end{equation}


\begin{thebibliography}{99}

\bibitem{FM2} J. Fr\"ohlich, P.A. Marchetti, Commun. Math. Phys. {\bf
112},343 (1987)
\bibitem{Wils} K.G. Wilson , Phys. Rev. {\bf D 10}, 2445 (1974)
\bibitem{Vill} J. Villain, Jour. de Phys. {\bf  36}, 581 (1975)
\bibitem{FM1} J. Fr\"ohlich, P.A. Marchetti, Europhys. Lett {\bf 2}, 933 (1986)
\bibitem{OS1} K. Osterwalder, R. Schrader, Commun. Math. Phys. {\bf 31},
33 (1975)
\bibitem{OS2} K. Osterwalder, R. Schrader, Commun. Math. Phys. {\bf 42},
281 (1975)
\bibitem{thooft} G. 't Hooft, Nucl. Phys. {\bf B 138}, 1 (1978)
\bibitem{pisa1} L. Del Debbio, A. Di Giacomo, G. Paffuti,
Phys. Lett. {\bf B 349}, 513 (1995)
\bibitem{Guth} A. Guth, Phys. Rev. {\bf D 21}, 2291 (1980)
\bibitem{KennKing} T. Kennedy, C. King, Commun. Math. Phys. {\bf 104}, 327
(1986)
\bibitem{FS1} J. Fr\"ohlich, T. Spencer, Commun. Math. Phys. {\bf 81}
,527 (1981)
\bibitem{FS2} J. Fr\"ohlich, T. Spencer, Commun. Math. Phys. {\bf 83}
,411 (1982)
\bibitem{Bryd} D. Brydges , {\em A short course in cluster expansions}, in :
Proceedings of 1984  Les Houches summer school, K. Osterwalder and R. Stora
editors (North-Holland)
\bibitem{Seil} E. Seiler, {\em Gauge theories as a problem of
 constructive quantum field theory and statistical mechanics}. Lecture
 Notes in Physics, Vol. 159 (Springer 1982)
\bibitem{Grad} I.S. Gradshteyn, I.M. Ryzhik: eq. 8.445 in {\em Table of
 integrals,series and products} (Academic Press)
\bibitem{tesi} P.A. Marchetti, PhD thesis: {\em An euclidean approach to
	 the construction and the analysis of the soliton sectors}
	(SISSA, 1986)
\bibitem{pisa2} A. Di Giacomo, G. Paffuti, preprint {\bf IFUP-TH/28-97}:
submitted to Phys.Rev {\bf D}
\end{thebibliography}
\end{document}